\documentclass[11pt,a4paper]{article}

\usepackage[usenames,dvipsnames]{xcolor}
\usepackage{colortbl}
\usepackage{multirow}

\usepackage{titling}

\usepackage[T1]{fontenc}
\usepackage[utf8]{inputenc}
\usepackage{textcomp} 

\usepackage{amssymb,amsfonts,amsmath}
\usepackage{hyperref}
\usepackage{url}

\usepackage{authblk}

\usepackage{booktabs,tabularx,ltablex,longtable}
\usepackage{eucal} 
\usepackage{paralist}
\usepackage{enumitem}

\usepackage{footnote}
\usepackage{graphicx}
\usepackage[left=1.5cm, right=1.5cm]{geometry}

\usepackage[style=mla,sorting=nyt,backend=bibtex,%
            noremoteinfo=false,mincitenames=1,maxcitenames=2]{biblatex}
\newcommand{\etal}{\emph{et al.}}
\newcommand{\ie}{\emph{i.e.,}}
\newcommand{\eg}{\emph{e.g.,}}

\newcommand{\kshell}{$k$-shell}
\newcommand{\kcore}{$k$-core}
\newcommand{\ks}{k_\text{s}}

\newcommand{\mycb}{\textsc{CB}}
\newcommand{\myev}{\textsc{EV}}
\newcommand{\myrel}{\textsc{REL}}
\newcommand{\mycor}{\textsc{COR}}
\newcommand{\mypub}{\textsc{PUB}}
\newcommand{\myall}{\textsc{ALL}}

\DeclareAutoCiteCommand{footnote}[f]{\footcite}{\footcites}

\graphicspath{{graphics/}}

\title{Quantifying Women's Marginalisation in Ibero-American Film Culture During the First Half of the Twentieth Century: A Network-Science Proposal\thanks{This research is funded by the ERC StG project ``Social Networks of the Past: Mapping Hispanic and Lusophone Literary Modernity,  1898-1959'' (Grant agreement No. 803860).}}

\author[$\dagger$,$\ddag$,$\star$]{Ainamar Clariana-Rodagut}
\author[$\dagger$,$\star$]{Alessio Cardillo}

\affil[$\dagger$]{Universitat Oberta de Catalunya -- Internet interdisciplinarily Institute (IN3), Barcelona, Spain}
\affil[$\ddag$]{Philipps-Universit\"at Marburg, Germany}
\affil[$\star$]{aclariana@uoc.edu; acardillo@uoc.edu}

\date{}

\begin{document}

\begin{titlepage}

\maketitle

\begin{abstract}
The research presented here uses the tools of social network analysis to empirically show a socio-cultural phenomenon already addressed by the social sciences and history: the historical marginalisation of women in the field of cinema. The novelty of our approach lies in the use of a large amount of heterogeneous historical data. On the one hand, we built a network of interactions between people involved in the film field in Ibero-America during the first half of the twentieth century. On the other hand, we propose a \kcore{} decomposition and a multi-layered analysis, as a quantitative way to study the position of women within the cultural melieu. After conducting our analysis, we concluded that women were mostly situated in the outer \kshell{s} of the empirical network, and their distribution was not uniform across the \kshell{s}. From a qualitative perspective, these results can be interpreted as the consequence of the lack of evidence of the participation of women in the public sphere.
\end{abstract}
 
\end{titlepage}

\section{Introduction}

The digital humanities blossomed as a discipline in the early twenty-first century. Since then, a multitude of computational-science methods from the computational sciences have been imported to the humanities. Digital tools have been successfully used in a plethora of disciplines, including history, philosophy, philology, and cultural studies, to analyse large amounts of data that would be impossible to manage using traditional tools. This has allowed for the discovery of new patterns and unidentified phenomena \autocites[2014]{schich-2014}[2018]{fraiberger-2018}[2013]{brughmans-jour_arch-2013}[2017]{fulminante-front_dig_hum-2017}. Nonetheless -- to the best of our knowledge -- network analysis has barely been applied to the study of film history. The research presented here seeks to fill this knowledge gap and uses the tools of social network analysis to empirically explore a socio-cultural phenomenon that has already been studied in the social sciences and history, namely, the historical marginalisation of women in the cinema field. Yet, the novelty of our approach lies in the use of a large amount of heterogeneous, historical data and in our proposed method to analyse the position of women within social networks linked to the cultural field. This approach allows us to highlight certain yet-unknown nuances of marginalisation and tie them to the network's structural properties and the nature of the connections composing it.

Our analysis has two goals. On the one hand, we seek to detect the social position of women in the cultural medium, specifically regarding film, in Ibero-America over the first half of the twentieth century, as per the data compiled in our database. To do so, we apply social network analysis and use a gender perspective and a relational approach. On the other hand, we propose using \kcore{} decomposition and multi-layered analysis as a quantitative way of detecting the marginal position of women in the cultural milieu.

The data we rely on for our analysis comes from a dataset comprising data on women who participated in the Ibero-American film field during the first half of the twentieth century. This dataset was extracted from the database created by the participants of the ERC project ``Social
Networks of the Past: Mapping Hispanic and Lusophone Literary Modernity, 1898-1959'' and members of the \href{http://globals.research.uoc.edu/}{{GlobaLS}} research group. Our dataset was built as part of the cinema and the women-focused subprojects of the larger, aforementioned ERC project.

The methodological proposal and analysis undertaken for this paper are the result of a joint effort between Alessio Cardillo -- a complex systems scientist -- and Ainamar Clariana -- a humanist deploying the gender perspective who specialises in cinema studies. Interdisciplinarily, we seek to detect the influential women in our database. From a qualitative perspective, it is generally accepted that historiography has invisibilized women who worked in the film field during the silent and early cinema periods.\footnote{In our case, this lack of data is exacerbated by our geographical context of focus, which has been considered peripheral by film historians. See for example Richard Abel's \emph{Encyclopedia of Early Cinema} (2004), in which entries on filmmakers are organised according to the countries where they worked, with the number of filmmakers from the United States towering over other countries. Also, these data records contain far fewer women than men. For recent work recovering the stories of women in silent cinema, see the \href{https://wfpp.columbia.edu/}{{Women Film Pioneers project}}.} It means that the data we gathered for the proposed framework are scarce and sparse. Even so, we believe that %
\begin{inparaenum}[1)]
\item quantitative methods can contribute to the discussion around why women suffer marginalisation, and that
\item network science can also propose different ways to tackle the historical roles played by women in the cultural field.
\end{inparaenum}
This paper aims to empirically show how marginalisation conditioned them.

By visualising the roles and positions women had in the networks, we pose the following research questions. Firstly, were the women peripheral? After exploring this question and realising that women were effectively situated at the margins of our network -- even when deploying the gender perspective and positive discrimination in the data-gathering process, we ask, what was the quantitative effect of marginalisation? To this end, we use \kcore{} decomposition as a suitable method for quantitatively analysing the peripheral position of women in our networks.

This method allows us to see that women were mostly situated in the outer \kshell{s} of the empirical network, and their distribution was not uniform across the \kshell{s}. In contrast, the random counterpart networks display a more even distribution of women across the \kshell{s}. Therefore, we postulate that there was a \emph{social force} pushing women to the outskirts of the network. In light of our results, we consider that the marginalisation effect is not due to mere chance but, instead, to the way the cultural field is structured and behaves when analysed as a network. From a qualitative perspective, these results can be interpreted as the consequence of the lack of data on women. This does not imply in any way that women did not occupy key roles in the institutionalisation process that took place in the film field. A plausible argument regarding this scarcity of data hinges on women's disproportionately low occupation of positions in the public sphere, compared to their dense fabric of relations and collaborations in the private spheres. As the majority of data we rely on has been gathered from publications and official documents, they fail to grasp the exchanges that took place in private spheres \autocite[1998]{arendt-1998}. To address the data's scarcity, we have developed certain data-gathering strategies, such as building ego-networks around certain key women mediators and then paying special attention to their female collaborators. Alternatively, we may assign a category to relationships in our database to add women for whom we do not have information other than their relationship (including marriage) with a man who occupied a relevant position in the cultural field of that time. Using our historical knowledge, we also structured the data in order to highlight the names of women on whom no further meta-information was available. One strategy we followed when considering distinct types of interactions was to convert each type into a separate network, to retrieve all the women who had personal relationships with other well-known people who swayed great symbolic power in the film and cultural field. We made this decision considering that most women in relationships with a male artists or intellectuals would have probably carried out relevant activities too.

\section{Methodology}

\subsection*{Perspectives}

As this paper includes the word ``women'' in its title, we must reflect upon what we mean by this category. First of all, it is crucial to state that most of the women we are referring to were white and came from upper- and middle-class backgrounds. This, in fact, conditioned women's survival in the intellectual and cultural fields. Although, on the one hand, their condition as women inevitably limited their social roles, their position was, on the other hand, very privileged compared to that of other women who were not white or from the upper-middle class. Silvia Federici~\autocite[2004]{federici-2004} notes that as long as the sexual division of labour in the framework of analysis remained as powerful as it had been since the emergence of capitalism, the concept ``women'' remains a useful category for analysis. Federici's work has led us to realise the importance of women's strategies to legitimise their voices and work in the cinema field. One shared experience among most of the women in our database was their fight against the roles imposed on them. Their struggle for emancipation indicates that, for all of them, there in fact was a clear definition of what it meant to be a woman, a category they often referred to and described in their works. Given the cultural environment they were raised in, they were told that their primary duty as women was to marry and care for the other members of their families. In one way or another, most of the women we have data on fought against this imposition. The crux of their struggle involved avoiding the reproductive labour they were destined for, resulting in social rejection. Nevertheless, the very privileged environment they were part of helped these women boost their productivity in the artistic and cultural fields as a direct consequence of their rejection \autocite[2024]{clariana_rodagut-book_chap-2024}. In this sense, these women could occupy social positions that were previously barred from women, thanks to their newfound freedom.

Comparing men and women is useful as long as we use it to describe the differences themselves and their implications, but comparison proves useless if we assume that men and women operated under equal conditions and thus expect equal, or even comparable, results. In no way can the public roles of the women at hand be approached using the same methods we would use to analyse men's contributions within the same context. Although men's contributions have always been historically acknowledged, women's contributions did not receive the same praise because of their limited access to the public sphere. Therefore, if women's contributions were to be compared to their male counterparts' using the same tools of measurement, women would always appear less relevant -- or gifted -- in the history of culture. The above reasoning has led us to work on certain compensation strategies which, taken together, make up the method of analysis we propose in this paper.

\subsection*{Data}

Our analysis leverages our research group's data, which is stored in the \href{https://nodegoat.net/}{{Nodegoat}} Virtual Research Environment. The database has been filled and enriched since 2019 and, therefore, is the result of the collaboration between current and past members of our research group. We believe it is important to be clear about the work involved in compiling data, given that such labour is often invisibilized, as noted by Catherine D'Ignazio and Lauren Klein in their book \emph{Data Feminism} (2019)~\autocite{ignazio-2020}. This invisibility implies that the \emph{places} from which datasets are built are rarely acknowledged. Thus, little responsibility tends to be assumed around the biases affecting data-collection, cleaning, and systematisation processes. While knowledge should always be situated \autocite[1998]{haraway-1988}, it is imperative that we specify our places of enunciation as researchers. This means that we must take responsibility for our biases, recognize what is being left out and what we are taking for granted, and clarify where the data were gathered and what procedure was used to extract them. The goal of such a process is to avoid what Haraway calls ``the god trick,'' that is, the deployment of a perspective that is situated nowhere, presuming that what we are showing reflects reality, instead of reflecting just one of its parts, \emph{our version of it}. We aim to steer clear of the idea that the data cannot lie.

Thus, it is worth mentioning that the data-compiling team operates from the south of the Global North and that the team's salaries are paid by a prestigious funding institution, the European Research Council. The means available to conduct our research are greater than average. Most of the people who have taken part in gathering our data are cisgender and white, and come from middle-class backgrounds. The data are biassed by our own standpoint.

The project behind the data collection, ``Social Networks of the Past: Mapping Hispanic and Lusophone Literary Modernity (1989-1959),'' consists of four research lines. These lines investigate cultural assets in the Ibero-America of the first half of the twentieth century. We aim at finding the mediators who spearheaded the exchange of material (\eg{} books) and inmaterial (\eg{} ideas) goods, but also the circulation of people. We posit that these relationships and exchanges provide the best windows to understanding the contribution that Ibero-America made in the building of artistic modernity. Our main research objects are %
\begin{inparaenum}[1)]
\item translations in cultural and literary historical journals,
\item the International Committee of Intellectual Cooperation of the League of Nations,
\item film criticism in specialised and cultural historical journals, and
\item women of the first Ibero-American film clubs.
\end{inparaenum}
The information we gathered involves human and non-human agents and agencies \autocite[2005]{latour-2005}, and using the actor-network theory's tenet ``follow the actor,'' all the agents found during our source's harvesting phase were uploaded into our database to unveil the networks around our objects. Therefore, in our database, we can find data on associations (for example, linked to historical journals), cultural events, and people. We also have data on people with whom the participants of our main objects exchanged letters and who collaborated in books and articles published by our human actors. Given our data's intrinsic heterogeneity, our sources are varied, as they come from primary sources and secondary bibliography. Primary sources can be located in personal, municipal, and national archives, or in historical publications, thus including correspondence, historical documents, pictures, exhibition catalogues, member lists, etc.\footnote{These are some of the most relevant sources for each case study. For Mar\'ia Luz Morales's case study, we included data from \autocites[2017]{cabre-book-2017}[2015]{codina_canet-pres-2015}[2011]{fulcara_torroella-book-2011}[2018]{gonzalez_naranjo-online-2018}[2019]{morales_cabre-book-2019}[1990]{perez_villanueva-archive-1990}[2006]{real_mercadal-book-2006}[1993]{zulueta-book-1993}, as well as from the articles she published in newspapers and journals, and from newspaper reports published in \emph{La Vanguardia}. For Lola \'Alvarez Bravo's case study, we gathered data from Bravo \parencites*[1982]{alvarez_bravo-archive-1982}[1994]{alvarez_bravo-archive-1994}, Debroise \parencite*[2005]{debroise-book-2005}, Fuentes Rojas \parencite*[1995]{fuentes_rojas-book-1995}, Poniatowska \parencite*[1993]{poniatowska-book-1993}, and Rodr\'iguez-\'Alvarez \parencite*[2002]{rodriguez_alvarez-thesis-2002}. For the Victoria Ocampo case study, we included data from Artundo \parencite*[2008]{artundo-book-2008}, Leston \parencite*[2015]{leston-book-2015}, Liendo \parencite*[2017]{liendo-lirico-2017}, and Ocampo \autocites*[1935]{ocampo-testimonios-1935}[1941]{ocampo-testimonios-1941}[1957]{ocampo-testimonios-1957}[1980]{ocampo-correspondencia-1980}[1983]{ocampo-autobio-1983}. We have included data from Ocampo's correspondence published by the Houghton Library in the collection Victoria Ocampo papers. We also added the data published by Benedict \parencite*[2022]{benedict-data-2022} on Ocampo's publishing house and journal \emph{Sur}.}

One of the consequences of using heterogeneous sources is that we fed the records on the relationships based on participation in events. Yet, we do not possess data on associations' every event but, in most cases, we have data on just a few events organised by a subset of associations. Likewise, the information does not cover every actor at a specific event or association, but only those we managed to find in the sources we had access to. Most of the data in this paper has been manually gathered. During the data harvest and uploading, each member of the team added descriptors using well-established concepts from their own fields of expertise. For instance, we have included the descriptor ``cinema'' to describe a given event's topic, the specialisation of a journal, or the main topic of a specific text. These descriptors are written in English -- as this is our research group's communication language -- and in Spanish -- the main vehicular language used by the agents and agencies we work with.

Data heterogeneity also affects the types of relationships we establish between actors. The plurality of sources and people who have extracted/curated the data may also lead to inconsistencies, which we have tried to mitigate through constant verification. The metadata available on each human actor can also vary. In this sense, the \emph{invisibilization} of women in cultural history, and the consequent paucity of data on them, is worth mentioning. Therefore, regarding the human actors, we have considered meta-information on gender, name (first and last), and professional or participation-related ties to magazines, events, publications, and institutions, when available. We have also included personal relationships for the most prominent people in our database, that is, the ones we have been extensively working on from a qualitative point of view. These personal relationships include family and friends, as we consider this information relevant to our framework.

To address the lack of data on women, we have created a category on relationships (either personal or professional) in our database. Although we do not always possess the empirical evidence (\eg{} an article, photo, or list of collaborators) of the professional relationship between two agents, when working with women it is essential that we include the self-reflexive and personal texts in which these women declare their professional or personal relationship with certain relevant actors in the cultural or cinematographic fields.\footnote{See Anselmo \parencite*[2023]{anselmo-2023} and Amelie Hastie \parencite*[2007]{hastie-2007}.} In our category ``relationships,'' we have selected only personal, romantic relationships (\eg{} wife-husband, or unmarried couple) and those defined as professional relationships. Such a selection is based on our knowledge of the historical context. Our expertise, in fact, leads us to believe that women married to "prominent" men almost always worked in their shadows. See, for instance, the case of Lola \'Alvarez Bravo.\footnote{Other members of the Mexican group of artists surrounding Lola \'Alvarez Bravo include Graciela Amador Sandoval and Isabel Villase\~nor (respectively, David Alfaro Siqueiros's and Gabriel Fern\'andez Ledesma's wives) \parencite*[Cueva][2017]{cueva_tazzer-thes_psi-2017}. The latter -- according to Lola \'Alvarez Bravo -- died because her husband did not let her out of the house or let her be as creative as she needed \parencite*[Poniatowska][1993]{poniatowska-book-1993}. There was also Elena Garro (Octavio Paz's wife), whose husband forced her to burn her own writing due to his jealousy of her skills \autocite[2019]{garro-book-2019}. In the circle of intellectuals tied to the Lyceum club in Madrid we may find María Lej\'arraga, the ghostwriter for her husband, the well known Gregorio Martínez Sierra \parencite*[Liz\'arraga Vizcarra][2020]{lizarraga_vizcarra-book-2020}. For further references of women artists whose work was underrepresented by their husbands, refer to Grosenick \parencite*[2001]{grosenick-book-2001} or Birrell \parencite*[2021]{birrel-book-2021}.} Concerning professional relationships, we have traced them thanks to primary sources (\eg{} self-reflexive texts) or other sources (\eg{} photographs), providing empirical evidence of the existence of a relationship between two (or more) people. 

In sum, the main criterion to select relevant data within the whole dataset stored in Nodegoat was that the agents had to have some relationship with the film field. This criterion entails the selection of those institutions (journals, film clubs, federations, etc.), events, and people that played a role or were related to the film field, with the latter having worked or somehow participated in any such events, institutions, and journals. Our decision is based upon the paucity of data and the need to work with indirect data. Thus, even if for some people we only have data related to a single event, we will still consider them actors tied to the film field. In the next section, we provide a detailed description of how each channel (or type) of interaction has been defined and subsequently used to generate a network of people.

\subsection*{Channels of Interaction}

In mathematical terms, a \emph{graph} (or network) is an object made of a certain number of points (denoted by $N$) called \emph{nodes} or \emph{vertices}, connected by a given number of \emph{edges} or \emph{links} (whose number is equal to $E$) \autocite[2017]{latora-2017}. The number of connections of a node is called its \emph{degree}, and we denote it as $k$. An \emph{isolated} node is a node with no connections. We use the term \emph{path} to indicate the sequence of edges (or nodes) that needs to be crossed to go from the path's origin to its destination. A \emph{connected component} of a graph is the set of mutually reachable nodes via a path. The biggest component in the graph is called its \emph{giant component} \autocite[2017]{latora-2017}.

Below we present how each channel (\ie{} type) of interaction was used to build the corresponding network. Each of these networks can be thought of either as an independent entity or, alternatively, as a subnetwork (\ie{} a \emph{layer}) of a multiplex/multilayer network \autocite[2020]{bianconi-2020}.

\begin{description}
    \item[Collective Bodies]{Given the list of all the collective bodies available in our database, we:
    \begin{enumerate}
        \item Selected those collective bodies (\mycb{}) for which the field ``Kind of organisation'' contained one of the following terms: '\emph{Cine}', '\emph{cine}', '\emph{Ciné}', '\emph{ciné}', '\emph{Film}', and '\emph{film}'. This operation yielded all the \mycb{} classified as '\emph{Film Club}', '\emph{Film Archive}', '\emph{Venue}', and 'Film Distributor'.
        \item Selected those \mycb{} whose field ``Name'' contained one of the following terms: '\emph{Cine}', '\emph{cine}', '\emph{Ciné}', '\emph{ciné}', '\emph{Film}', and '\emph{film}'.
    \end{enumerate}
    These steps allowed us to select, overall, 422 \mycb{}. Then, for each \mycb{}, $i$, we extracted the set, $P_{\text{\mycb{}}} (i)$, of people involved with their activities but classified as prominent figures (\eg{} founders, organisers, managers, treasurers, etc.) or playing a relevant role (for instance, presenters, guests, authors, etc.) within them. Then, given the set $P_{\text{\mycb{}}} (i)$, we added an edge between each pair of elements. By considering only prominent figures, we ensured that the interactions we captured were those among people who were committed to the activities organised by the \mycb{} and therefore would have met at some point. That said, it is worth stressing that such an assumption implies that each \mycb{'s} prominent figure took part in all its activities (which might not always have been the case).
    } 
    \item[Events]{Given the list of all the events available in our database, we:
    \begin{enumerate}
        \item Selected those whose field ``Subject'' contains the word `Cinema'.
        \item Selected those whose field ``Name'' contained one of the following terms: 'cine', 'film', 'pelicula', and 'proyecci'.\footnote{The match is case insensitive.}
    \end{enumerate}
    These steps allowed us to select 99 events (\myev{}). Then, for each \myev{}, $i$, we extracted the set, $P_{\text{\myev{}}} ( i )$, of its participants, and added a connection between each possible pair of them. It is worth noting that some \myev{}'s participant list consists of only one -- or no -- person. When we only found one person, the node corresponding to the sole participant was added to the network without any edge connecting it.
    } 
    \item[Publications]{Given the complete list of all publications available in our database, we:
    \begin{enumerate}
        \item Selected those whose field ``Genre'' corresponded to `Film'.
        \item Selected those whose field ``Title'' contained either the word `film' or the word `cine'.\footnote{The match is case insensitive.}
        \item Selected those whose field ``Journal name'' contained one of the following terms: `film', and `cine'.\footnote{The match is case insensitive.}
    \end{enumerate}
    Using these criteria, we selected 64 publications (\mypub{}). From these \mypub{} we extracted the set of journals where they were published (a total of 16), and created a set, $P_{\text{\mypub{}}} (i)$, of people associated with each journal, $i$, either due to authorship or, for those related to cinema, due to the role they played within them. For the latter, we selected those whose role is one of the following: `Director', `Editor', `Co-director', `Editor in chief', `Co-founder', and `Founder'. This approach allows us to grasp not only authorship relationships but also -- and most importantly -- relationships between authors and editors. Indeed, many publications in specialised journals were submitted thanks to direct invitations from the editor, thus implying the existence of a relationship between editors and authors. We also assume that the people publishing pieces on cinema were interested in cinema as well.
    } 
    \item[Relationships]{Given the list of people involved in \mycb{}, \myev{}, or \mypub{} relationships, we extracted the list of (pairwise) personal relationships (\myrel{}) of the     type `Spouse', `Unmarried partner', and `Professional'. This step led us to include people not belonging to the \mycb{}, \myev{}, and \mypub{} relationships. Such a selection criterion translated into 601 connections. 
    } 
    \item[Epistolary Correspondence]{ Given the list of people involved in \mycb{}, \myev{}, \mypub{}, and \myrel{}, we searched through our epistolary database once again and added epistolary correspondence (\mycor{}) only if both the sender and the receiver belonged to the aforementioned list of people. We decided not to include all the correspondence from our database because we did not want to add new actors who were not connected to the cinema field to our network. Rather, we deemed only those actors with a role in the cinema field to have played a role in the articulation of the field in itself.
    } 
\end{description}
Table~\ref{tab:net_props} summarises the main structural features of these networks, comprising the number of nodes, $N$, of edges, $E$, the maximum number of connections per node, $k_{\max}$, the number of connected components, $N_{\text{comp}}$, (\ie{} pieces of the network in which it is possible to go from every node to every other node within it), the number of isolated nodes, $N_{\text{isol}}$, and the relative size of the giant component, $S$, (\ie{} the percentage of nodes belonging to the biggest connected component in the graph) \autocite[2017]{latora-2017}. We also computed the same indicators for the network merging together all the layers (\myall{}).

%
%
%
\begin{table}[h!]
    \centering
    \resizebox{0.7\linewidth}{!}{%
        \begin{tabular}{rlllllrl}
        \toprule
        \multicolumn{1}{c}{\textbf{Network}} & \multicolumn{1}{c}{$N$} & \multicolumn{1}{c}{$E$} & \multicolumn{1}{c}{$k_{\max}$} & \multicolumn{1}{c}{$N_{\text{comp}}$} & \multicolumn{1}{c}{$N_{\text{isol}}$} & \multicolumn{1}{c}{$S (\%)$} & \multicolumn{1}{c}{$D$}\\\toprule
        \myrel{} & \multirow{5}*{1367} & 6011 & 75 & 783 & 412 & 6.95 & 4\\
        \mycor{} &  & 101 & 91 & 1269 & 1268 & 7.24 & 2\\
        \mycb{} &  & 1315 & 23 & 1136 & 1101 & 1.76 & 20\\
        \mypub{} &  & 244 & 18 & 1323 & 1317 & 1.39 & 16\\
        \myev{} &  & 1898 & 95 & 1258 & 1250 & 7.02 & 58\\
        \rowcolor{black!30!white} \myall{} & 1367 & 4090 & 117 & 373 & 37 & 36.72 & 58\\
        \bottomrule
        \end{tabular}
    } 
    \caption{Structural features of the networks used in our study. See the main text for the details.}
    \label{tab:net_props}
\end{table}

\subsection*{Brief Introduction to Mathematical Concepts}

\subsubsection*{Jaccard Index}

Using different networks accounting for distinct channels of interaction raises the question of whether the connections grasped by each network are the same or not (and to what extent). One way to quantify the redundancy of the information encoded in two layers of a multiplex network is to compute the \emph{Jaccard index} of their edges' sets \autocite[2020]{bianconi-2020}. Given two generic sets $\alpha$ and $\beta$, their Jaccard index \autocite[1901]{jaccard-1901}, $J$, can be computed mathematically:
\begin{equation}
\label{eq:jaccard}
J(\alpha,\beta) = \frac{ \bigl| \alpha \cap \beta \bigr|}{\bigl| \alpha \cup \beta \bigr|}\, .
\end{equation}
The numerator $\bigl| \alpha \cap \beta \bigr|$ indicates the number of elements belonging both to sets $\alpha$ and $\beta$. The denominator $\bigl| \alpha \cup \beta \bigr|$, instead, denotes the number of elements belonging to either set $\alpha$ or $\beta$. The structure of Eq.~\eqref{eq:jaccard} implies that the values of $J(\alpha,\beta)$ can span from zero to one. A value of $J(\alpha,\beta) = 1$ indicates that the two sets have exactly the same elements (\ie{} one is an exact copy of the other), whereas $J(\alpha,\beta) = 0$ indicates that the two sets do not have any element in common (\ie{} they are completely different). A value of $J(\alpha,\beta)$ between zero and one indicates, instead, a certain degree of similarity between the two sets, with values of $J(\alpha,\beta)$ close to one denoting high similarity and those close to zero denoting low similarity.

\subsubsection*{\kcore{} Decomposition}

As we are interested in pinpointing those women who played a relevant role in cultural mediation in the Ibero-American film field of the first half of the twentieth century, we can leverage the network's formalism to identify them, pushing our analysis beyond dyadic interactions by considering mesoscopic structures -- \ie{} coherent structures made by groups of nodes. Amidst the plethora of possible mesoscopic structures \autocite[2019]{lambiotte-2019}, we decided to focus on a type of core-periphery structure called \emph{\kcore{} decomposition} \autocites[2000]{borgatti-2000}[1983]{seidman-1983}[2019]{kong-2019}. Such a technique has proven useful in a variety of domains \autocites[2007]{carmi-2007}[2018]{morone-2018}[2019]{kong-2019}, but one in particular is especially appealing for our case. It turns out that \kcore{} decomposition identifies the most influential spreaders in a network more efficiently than by ranking them according to, for instance, their number of connections \autocite[2010]{kitsak-2010}. Cultural mediators, like the women we are working on, can be taken, to a certain extent, as spreaders in the sense that they played the key role of mediating between spheres and cultures. In this sense we considered the ability to spread (as defined by network science) as similar to mediation (as defined by the social sciences) \autocite[2018]{roig_sanz-2018}. This underpins our choice of using \kcore{} decomposition to pinpoint women cultural mediators.

The \kcore{} decomposition of a network is an iterative pruning process decomposing it into a set of concentric (\ie{} onionlike) \emph{\kshell{s}}. The algorithm consists in recursively removing the nodes having less than $k$ connections (\ie{} first $k = 1$, then $k = 2$, and so on).\footnote{These nodes could originally have more than $k$ connections, but the removal of other nodes connected to them decreases their effective number of connections.} Under this assumption, a \kshell{} is the set of nodes belonging to the $k$-th core but not to the ($k + 1$)-th core (that is why the structure is akin to that of an onion). The \kshell{} index, $\ks$, of the deepest \kshell{} available for graph $G$ (\ie{} its maximum on the graph $\max_{G} ( \ks )$) is its \emph{degeneracy}, $D$.

Given the aforementioned properties of the \kcore{} decomposition, we studied the fraction of nodes of each gender in each \kshell{}. To ensure that our observations are not the result of mere chance, we compared them to their analogue, measured in a randomised version of the network.

\subsubsection*{Graph Randomisation}

Given a network $G$ with $N$ nodes and $E$ edges, $G(N,E)$, the randomisation process generates a network $G^\prime (N,E)$ which, besides the number of nodes and edges, also preserves each node's number of connections. Such a random counterpart can be obtained from network $G$ using the edge-swapping rewiring mechanism proposed by \autocite[2018]{fosdick-2018}, developed from the \emph{configuration method} \autocite[1995]{molloy-1995}. In a nutshell, the edge-swapping method selects uniformly at random two edges, $(a,b)$ and $(c,d)$, of the network $G$, and replaces them -- if they do not exist already, or generate self-loops -- with either pair $(a,c)$, $(b,d)$ or, alternatively, with the pair $(a,d)$, $(b,c)$. Both replacements ensure the conservation of each node's degree. For this study, we repeated the edge-swap step several times to ensure that the final network $G^\prime (N,E)$ was sufficiently distinct from the original one. We repeated the whole randomisation procedure several times (in our case, 50) to ensure the statistical significance of our results.

\section{Results}

\subsection*{Quantifying Women's Marginalisation Using \kcore{} Decomposition}

One way of quantifying the marginalisation of women in the film field is to study their presence -- and abundance -- across the network's \kshell{s}, as outer ones tend to be populated mainly by nodes belonging to the network's outskirts, whereas deeper \kshell{s} are populated by more influential spreaders, which also tend to occupy central positions in the network. To this end, we have built a network -- labelled as \myall{} -- obtained by collapsing together all the layers (\ie{} the \myrel{}, \mycor{}, \mycb{}, \mypub{}, and \myev{} networks) of the multiplex network. The collapse of distinct layers into a single network implies that the presence of an edge in the \myall{} network stems from its existence in, at least, one of the layers. Table~\ref{tab:net_props} summarises the structural features of the \myall{} network. Upon inspection, we may highlight that the number of edges, $E$, and the size of the biggest connected component, $S$, are considerably larger than those of the single layers. These differences, combined with the smaller number of connected components, $N_{\text{comp}}$, and isolated nodes, $N_{\text{isol}}$, tell us that the \myall{} network is considerably more cohesive than the layers' networks when taken separately. With this picture in mind, we can study the \kcore{} decomposition of the \myall{} network. Specifically, for each \kshell{}, we have counted its number of nodes classified either as men, women or ``unknown''\footnote{We use this category to indicate the lack of metadata on the gender of a person.} and divided such numbers by the total number of nodes in the \kshell{}.\footnote{This procedure ensures that the sum of the fractions of men, women, and unknown nodes of a given \kshell{} is always equal to one.}

\begin{figure}[ht]
    \centering
    \includegraphics[width=0.95\textwidth]{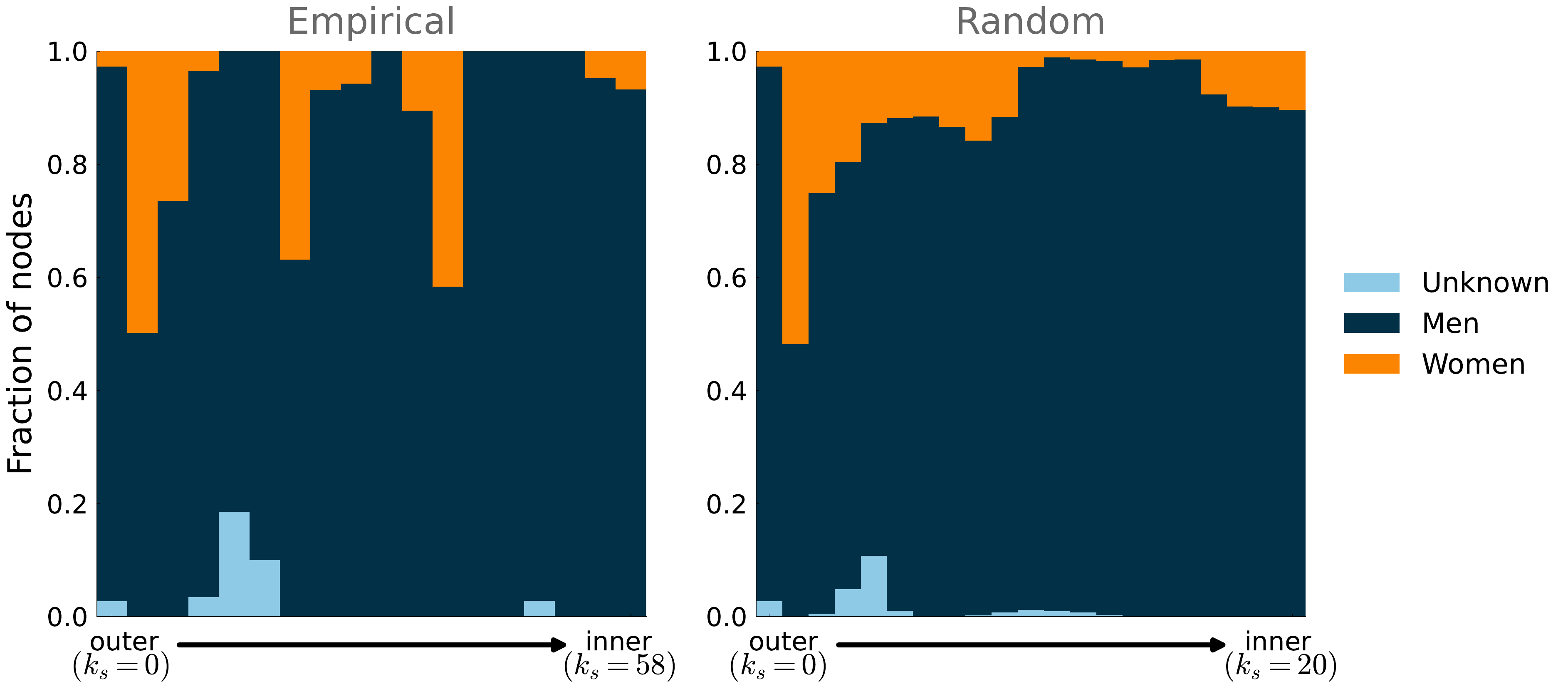}
    \caption{Organization of the nodes grouped by gender across the \kshell{s}. The horizontal axis indicates the \kshell{} index, $\ks$, whereas the vertical axis denotes the fraction of nodes with a given gender in each \kshell{} (inner \kshell{s} are the most central ones). The left panel displays the results for the empirical \myall{} network, whereas the right panel displays the case of the random counterpart (results are averaged over 50 realisations of the rewiring process).}
    \label{fig:kcore-comparison-emp_rand-all}
\end{figure}

The left panel of Fig.~\ref{fig:kcore-comparison-emp_rand-all} displays the fraction of women, men, and unknown people present in each \kshell{} of the empirical network. Notably, the women tend to concentrate in a few \kshell{s}, especially in the outer -- and shallower -- ones. However, the same analysis performed on the randomised networks delivers a distinct picture. By glancing at the right panel of Fig.~\ref{fig:kcore-comparison-emp_rand-all}, we notice how women's presence tends to be more widespread across all the \kshell{s}. The fraction of women appearing in a \kshell{} appears more homogeneous in the random counterpart than in the empirical network. These features make us suspect that there could be some kind of ``force'' exacerbating women's marginalisation by pushing them towards \kshell{s} where their influence as cultural mediators is weaker.

%
%
%

\subsection*{Decoupling the Source of Marginalisation}

The marginalisation of women within the \kcore{} decomposition observed in the \myall{} network (displayed in Fig.~\ref{fig:kcore-comparison-emp_rand-all}) raises the question of whether such a phenomenon stems directly from a single -- or a few -- channels of interaction or is due to their interplay. It has been proved that the structural features of networks obtained by merging together the layers of a multiplex network can be quite different from those of their layers considered singularly \autocite[2013]{cardillo-2013}. To find whether marginalisation is directly related to the multilayered nature of our system, we leveraged the full structure of the multiplex network and repeated our analysis of each of the layers separately. Such a procedure allows us to understand how marginalisation affected women in different facets of their activities (\ie{} channels of interaction). We hypothesise that each channel of interaction would show different levels of marginalisation depending on the type of interaction involved: more private channel of interaction (such as \mycor{} or \myrel{}) would display smaller levels of marginalisation among women, whereas a more public channel of interaction (\eg{} \myev{}) would reveal higher levels of marginalisation.

\begin{figure}[ht]
    \centering
    \includegraphics[width=0.55\textwidth]{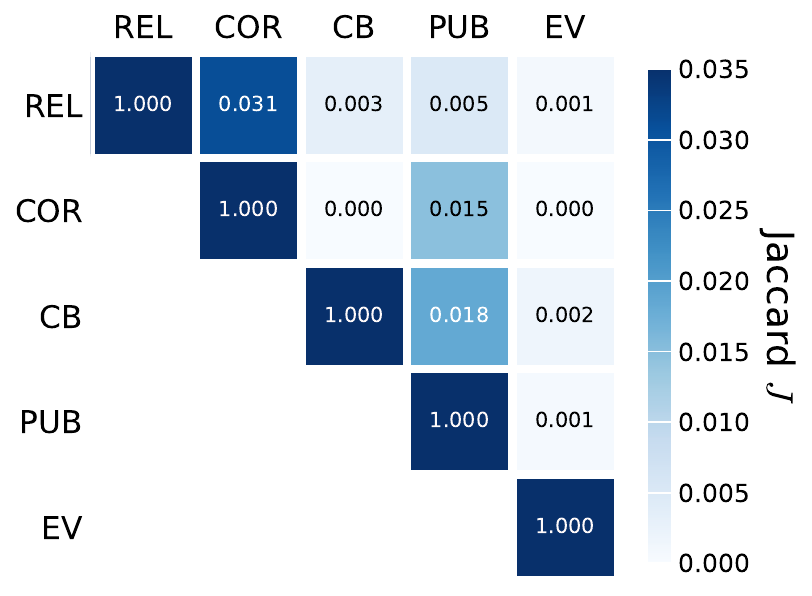}
    \caption{Heatmap matrix encoding the values of the Jaccard score, $J(\alpha,\beta)$, computed between pairs of networks $\alpha$ and $\beta$ corresponding to the layers of the multiplex network. Light shades denote small similarities, whereas those on the diagonal are by definition equal to one. The symmetry of $J(\alpha,\beta)$ allows us to display only the matrix's upper half.}
    \label{fig:jaccard-matrix-layers}
\end{figure}

To quantify the overlap of information in each layer pair, we use Eq.~\eqref{eq:jaccard} to compute the pairwise Jaccard index $J(\alpha,\beta)$ between layers $\alpha$ and $\beta$. Figure~\ref{fig:jaccard-matrix-layers} shows the heatmap matrix of the values of $J$ computed for all the pairs of layers. Except for the elements of the main diagonal (comparing a layer with itself, thus translating to a value of $J = 1$) all the remaining values of $J$ are quite small. This means that the amount of redundant (\ie{} overlapping) information encoded in our layer networks is quite small, ranging between 0.0 and 0.031. In other words, our layer networks are quite distinct from one another. Such differences justify the use of multiplex network formalism \autocite[2020]{bianconi-2020}.

%
%
%

\begin{figure}[ht]
    \centering
    \includegraphics[width=0.8\textwidth]{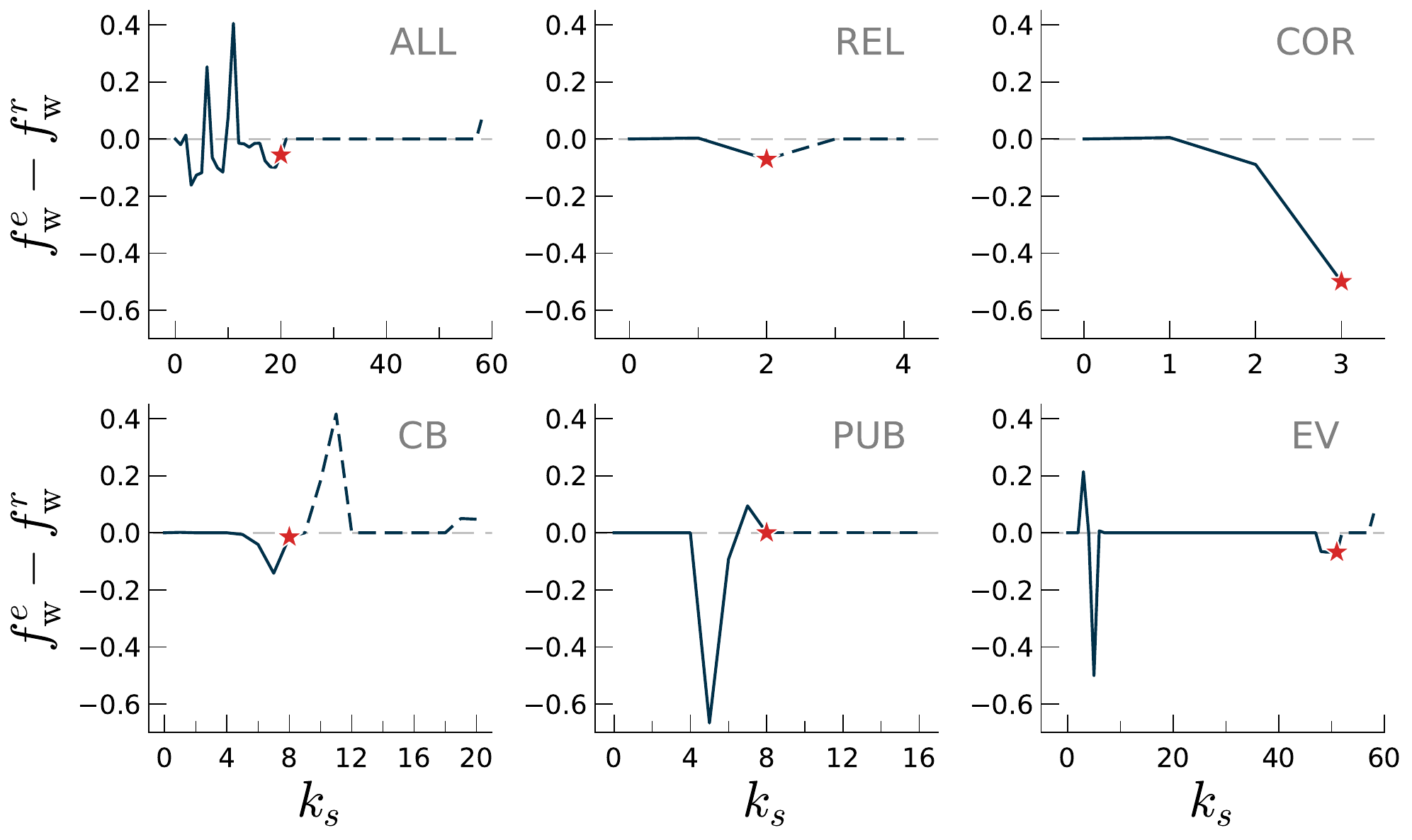}
    \caption{Differences between the fraction of women nodes belonging to \kshell{}, $\ks$, of the empirical network, $f_{W}^{e}$, and its average counterpart computed in a randomly rewired network, $f_{W}^{r}$. Each panel accounts for a distinct network, whether the whole network (\myall{}) or a single layer (\myrel{}, \mycor{}, \mycb{}, \mypub{}, and \myev{}). The red star indicates the deepest \kshell{} (\ie{} the network's degeneracy ($D$)) obtained in the random network counterpart $D^{r}$, whereas the dashed line denotes the \kshell{s} for which only the empirical value, $f_{W}^{e}$, exists. Random values are obtained by averaging the results over 50 realisations.}
    \label{fig:kcore-comparison-emp_rand-layers}
\end{figure}

To delve into the source of women's marginalisation, for each of the layer networks, including the \myall{} one, we computed the difference between the empirical fraction of women belonging to a given \kshell{}, $f_{W}^{e}$, and the same quantity computed in the randomised networks, $f_{W}^{r}$ (see Figure~\ref{fig:kcore-comparison-emp_rand-layers}). Specifically, for each network $G$ we computed the difference $\Delta f_{W} = f_{W}^{e} - f_{W}^{r}$ for each \kshell{} available. A positive difference (\ie{} $\Delta f_{W} > 0$) indicates that for that \kshell{}, there are (proportionately) more women in the empirical network than in the random counterpart. Conversely, a negative difference (\ie{} $\Delta f_{W} < 0$) indicates that -- on average -- the random counterpart has proportionately more women than the empirical one. Finally, a difference equal to zero (\ie{} $\Delta f_{W} = 0$) indicates that the fraction of women in that \kshell{} is the same in both the empirical and the random networks.

%
%
%

By inspecting Figure~\ref{fig:kcore-comparison-emp_rand-layers}, we may highlight a few remarkable features. First, random networks tend to have a shallower \kshell{} structure than their empirical counterparts. The red star indicates the position of the value of the degeneracy for random networks, $D^{r}$. We notice how for all networks (except the \mycor{} one), $D^{r}$ is smaller than in the empirical case (see also Table~\ref{tab:net_props}). Such a recoil has been observed already in empirical and artificial networks \autocite[2020]{malvestio-2020} and is due to the shattering of tightly knit groups of nodes (\eg{} communities) operated by the random rewiring mechanism. Another feature of Fig.~\ref{fig:kcore-comparison-emp_rand-layers} is  that marginalisation occurs mainly in the \myev{} layer but also in the \mypub{} and \mycb{} layers. In contrast, it is absent in those layers corresponding to ``private'' interaction channels, such as \myrel{} or \mycor{}. These opposite behaviours originate from the intrinsic nature of the \myev{}, \mypub{}, and \mycb{} networks, which proceed from the projection of two-mode networks (\ie{} networks with two types of nodes with connections existing only between nodes of different type) into single-mode networks rich with cliques (\ie{} groups of nodes all connected with each other). This dichotomy suggests that public interactions were more prone to marginalise women than those occurring in the private sphere. This is why it is crucial that we compare the single-mode networks' results (\mycor{} and \myrel{}) to those of two-mode networks (\myev{}, \mypub{}, and \mycb{}) and give them the same value when analysing women's contributions to the cultural field. Finally, we observe how the collapse of the available interactions' channels into a single layer gives rise to a less defined marginalisation scenario than when the layers' networks are studied independently.

\section{Conclusions}

The methods proposed in this manuscript can help us quantitatively understand the marginalisation of women in the Ibero-American film field of the first half of the twentieth century. The novel and valuable nature of this joint research effort does not lie in a discovery but, rather, in a methodological proposal combining quantitative tools of social network analysis with a feminist perspective in the field of historical cultural phenomena. A growing number of studies are analysing data from several cultural and social fields through the lenses of gender and social networks analysis \autocites[2010]{smith_doerr-2010}[2015]{lutter-2015}[2020]{morgan-2021}[2022]{macedo-2022}[2022]{wapman-2022}[2023]{herrera_guzman-2023}. Data science has also been applied to the study of the history of cinema, especially the history of reception, exhibition, and distribution practices \autocite[2011]{maltby-book-2011}. Nontheless, only a few works apply social network analysis tools to the study of cinema history, though they only use contemporary data \autocites[2020]{verhoeven-2020}[2023]{loist-eur_j_med_stud-2023}. In this sense, the exceptionality of our research lies in the combination of applying social network analysis methods to historical data from cinema and cultural history while deploying the gender perspective.

One of the main results of our work is the empirical confirmation of existing qualitative knowledge, namely, that during the first decades of the twentieth century, women were rather marginalised from the cultural public sphere. This phenomenon is easily traceable in the literary and publishing world \autocites[1995]{king-1995}[2009]{kowaleski_wallace-2009}. However, the same conclusion holds for cinema history in the same period, as publications and the creation of theoretical film knowledge through publications were particularly relevant both in the emergence of film cultures \autocite[2017]{navitski-2017}, as well as in the process of institutionalisation of the film field \autocite[1979]{bourdieu-1979}. Yet, the theoretical and historical  knowledge generated in spaces in which women often participated collectively, such as film clubs, ended up being attributed to individual, usually male, authors. Likewise, the events organised around the medium -- like competitions and festivals -- and their ``visible figures'' would stand as cornerstones in the institutionalisation of a field that was not yet professionalised. As Mary Beard \parencite*[2017]{beard-2017} pointed out in her classic text, certain mechanisms in Western culture, from its foundation, have hindered and continue to hinder women's access to the spheres of power. This means that a big part of the cultural activity women performed occurred in the private spheres.

After using affirmative-action strategies in the data-collection process, as well as in the data's systematisation and analysis, our quantitative analysis has bolstered this idea. The first strategy was to use the Latourian method for building networks and to trace the relationships starting with historical female agents. After that, we used the Latourian method of ``follow the actor,'' consisting in choosing case studies around which to build networks. We selected the case studies after qualitative historical research and now, after qualitative analysis, we have the certainty that those cultural mediators were relevant within the historical period under analysis. Despite the differences between the actor-network and the networks from graph theory, certain similarities allow us to use graphs to study social phenomena, as proposed by Venturini \etal{} \parencite*[2019]{venturini-2019}. Here, we go one step further by leveraging tools from both approaches to reach our goal: to analyse the marginalisation of women in a specific historical-cultural context.

Another strategy we used to trace women in the data-structuring process was to add the category ``relationships.'' The main purpose of this strategy was to introduce information on personal and work relationships mentioned in underexplored historical sources, whose legitimacy has been neglected until now. After conducting qualitative research within our framework, we discovered that most of the women who were married or who were lovers of men playing a prominent role in the cultural fields carried out relevant cultural activities as well.\footnote{For instance, the work of Lola \'Alvarez Bravo was considered irrelevant for Mexican culture until the '80s, the decade of her death. The most well-known examples in Spain include \emph{las Sinsombrero}, who adopted prominent roles in the cultur and intellectual fields of the Spanish pre-Civil War period, whose contributions had not been acknowledged until very recently \autocites[2005]{alonso_valero-pandora-2005}[2016]{ballo-book-2016}. These women were either close friends, lovers, or spouses of the intellectuals and male artists of the \emph{Generación del 27}, or were close to the \emph{Residencia de Estudiantes}. While the male artists and  intellectuals close to these groups have been prominently studied by  cultural historians for a long time, women who also took part in the same cultural fields have been completely neglected. We might mention Teresa de León, Concha Méndez, and María Luisa Muñoz de Vargas (spouses of Rafael Alberti, Manuel Altoaguirre, and Rogelio Buendía, respectively), or Maruja Mallo and Margarita Manso (close friends of Salvador Dalí and Federico García Lorca). The same pattern can be spotted in similar frameworks. Women who were surrealist artists and intellectuals were invisibilised by cultural historians and overshadowed by their male  counterparts, including Leonora Carrington and Dora Maar (respectively, Max Ernst and Picasso's lovers).} Furthermore, some evidence, such as photographs from the period, show that the participation of women in spaces considered eminently masculine in historiography was actually quite high.\footnote{Such as the first Western film clubs \autocite[forth.]{clariana-rodagut-forth}.} However, as a consequence of the structural chauvinism affecting historiography, some evidence had not been fully considered in the histories of Ibero-American cinema.

Despite the aforementioned compensation strategies, the data used to carry out our study remains limited and very heterogeneous. The former, as expected, stems from the little attention these women have received from historiography. This is relevant because only by scraping data from different fields and disciplines can we prove the relevance of the cultural mediators we are studying \autocite[2018]{roig_sanz-2018}. We decided to use the \emph{multilayer network} formalism and combine it with \kcore{} decomposition to address the data's heterogeneity. Although we do not have access to information on every woman's connection , we still had key information assisting us in investigating the roles women played in the network. For example, we possessed data on relevant connections, or on key events or organisations they took part in. Their strategic positions, allowing us to label them as cultural mediators, were as important -- or more so -- as the number of connections they had. If we would have analysed their relevance looking exclusively at their number of connections, they would have appeared much less relevant to cultural history than their male counterparts. In contrast, by considering their mediating roles, we have assessed their contributions to the cultural field in a fairer way. In network analysis, a proxy to measure mediating abilities is to measure the ability to diffuse information efficiently \autocite[2010]{kitsak-2010} and keep the system united as a whole \autocites[2007]{carmi-2007}[2019]{kong-2019}, thus allowing us to use the \kcore{} decomposition method.

Our choice to work with a multi-layered network is based on two factors: first, it has preserved all the channels of interaction that we deemed relevant, keeping us from losing any of the mediating facets or roles that these women played in the cultural field. Second, it has allowed us to see -- at a later stage -- the differences in the way each layer of interaction is structured. We allude to the \emph{cultural field} because the non-institutionalisation of the film field often implied that cultural activities related to film were organised alongside other literary, artistic, and intellectual activities. Therefore, although the data concerns the field of cinema, their relationship with the latter may often have been indirect. For instance, when including the data from a film critique (namely the author/s' name/s), we also added the names of people involved not only with the volume in which the critique was published, but also with the journal in which the piece was published. In this sense, the use of multi-layer network formalism allowed us to preserve information about the relationships that women established through different channels, such as publications, involvement in institutions, or participation in events.

Then, we performed \kcore{} decomposition on all the layers separately, as well as on the post-collapsation single layer. Our intuition was that, given the nature of our data and our object of study, different interaction layers would probably display different behaviours, as we saw in Figure~\ref{fig:kcore-comparison-emp_rand-layers}. In addition, the intensity of the marginalisation effect was amplified in the collapsed network, though this was not the case when each layer was analysed individually. In fact, this effect did not occur in certain layers (\eg{} the \mycor{} network). When performing \kcore{} decomposition on separate layers (networks), we observed that the marginalisation effect is stronger in the \myev{} network. In this layer, women fall into the outer \kshell{s} more often than in the random network counterpart. The strength of the marginalisation effect is stronger in the \myev{} layer because, by definition, the network corresponding to one event is a \emph{complete graph} or a \emph{clique} (\ie{} a graph where each node is connected to every other node). The edge-swap stage adopted by our randomisation process destroys such cliques, thus making the whole network's \kcore{} structure shallower \autocite[2020]{malvestio-2020}. A similar reasoning applies to the \mypub{} and \mycb{} networks, albeit to a much smaller extent.

Another possible explanation for the marginalisation phenomenon concerns the nature of the data. Despite our mitigation strategies, our data for public activities like events, come from publicly available information (mostly newspapers and magazines). The press releases advertising or reviewing events (our information's source) were less keen to include women in lists of participants, thus making their appearance more marginal than men's. Finally, another explanation for the marginalisation effect has to do with the features of people's collaboration networks. As the majority of unacknowledged women in our dataset come from associations' lists of participants, they form a sort of ``private club,'' inflating the amount of woman-to-woman connections \autocite[2002]{newman-2002}.\footnote{This phenomenon is called \emph{assortative mixing} and has been observed in several types of networks, including social ones (see \cite[2020]{verhoeven-2020} and references therein).} However, such a correlation gets weaker in the randomised network.

Despite the appeal and advantages of using a quantitative approach to complement qualitative and historical insight, there are some caveats. For instance, \kcore{} decomposition assumes that the network under scrutiny could be made of several components, but their structure should be complex enough to ensure a moderate value of degeneracy. This means that the insight provided by the \kcore{} decomposition is directly proportional to the total number of \kshell{s} available. For example, the \kcore{} decomposition of a network comprised of a group of disjoint yet complete graphs (\eg{} when many events do not have any participant in common) yields nothing other than the members of these graphs (\ie{} the event's participants) grouped by size (\ie{} number of attendants). This implies that nodes less entwined with the ``bulk'' of the system will be relegated to the outer \kshell{s} of the network, even when frequently interacting with smaller groups. with marginalised minorities -- like women in film culture -- which only have scant bonds with the rest of the system, might decrease the amount of information extractable via the \kcore{} decomposition. Another shortcoming stems from cliques' resilience to the \kcore{'s} iterative pruning process. Such resistance makes nodes belonging to these cliques (events) fall into inner \kshell{s}, thus pushing nodes (people) belonging to larger cliques (\ie{} attending events with many participants) into deeper \kshell{s}, even though these cliques do not share any, or no more than a few, nodes with each other.

Given these limitations, one possible workaround could be to cross-examine the mesoscopic structure of the network and extract its \emph{community structure} \autocite[2019]{lambiotte-2019} to study how the latter overlaps with the innermost \kshell{s} \autocite[2020]{malvestio-2020}. For instance, some of the events organised by the Lyceum Club might translate into specific communities falling to the innermost \kshell{s}.

Despite the limitations of the methodology proposed in this manuscript, we believe that our attempt is valuable to the study of minorities, like women, and the social spaces they occupy. While our methodological proposal can be used in a contemporary context, it has been developed specifically for the study of historical data concerning a minority group. In sum, we believe that collaboration between specialists of different expertise and perspectives harnesses a great potential that has yet to be discovered, which could revolutionise the field of the humanities and pave the way for new and exciting challenges in network science.


\subsection*{Acknowledgements}

The authors would like to thank Vensislav Ikoff for his support in extracting and curating the data, as well as Diana Roig-Sanz and Malte Hagener for their suggestions.\newline\newline
The numerical analysis has been carried out using the NumPy and NetworkX Python packages \autocites[2011]{van_der_walt-2011}[2008]{hagberg-2008}. Graphs have been prepared using the Matplotlib Python package \autocite[2017]{hunter-2007}.


\nocite{*}
\printbibliography

\end{document}